# THE PERFORMANCE OF CONVOLUTIONAL CODING BASED COOPERATIVE COMMUNICATION: RELAY POSITION AND POWER ALLOCATION ANALYSIS

Cebrail ÇIFLIKLI[1], Waeal AL-OBAIDI[2] and Musaab AL-OBAIDI[2]

[1]Electronic Tech. Program/ Electronics and Automation,
Erciyes University, Kayseri, Turkey
[2]Faculty of Electrical and Electronic Engineering,
Erciyes University, Kayseri, Turkey

*ABSTRACT*

*Wireless communication faces adversities due to noise, fading, and path loss. Multiple-Input Multiple-Output (MIMO) systems are used to overcome individual fading effect by employing transmit diversity. Duo to user single-antenna, Cooperation between at least two users is able to provide spatial diversity. This paper presents the evaluation of the performances of the Amplify and Forward (AF) cooperative system for different relay positions using several network topologies over Rayleigh and Rician fading channel. Furthermore, we present the performances of AF cooperative system with various power allocation. The results show that cooperative communication with convolutional coding shows an outperformance compared to the non-convolutional, which is a promising solution for high data-rate networks such as (WSN), Ad hoc, (IoT), and even mobile networks. When topologies are compared, the simulation shows that, linear topology offers the best BER performance, in contrast when the relay acts as source and the source take the relay place, the analysis result shows that, equilateral triangle topology has the best BER performance and stability, and the system performance with inter-user Rician fading channel is better than the performance of the system with inter-user Rayleigh fading channel.*

## 1. INTRODUCTION

The transfer of information through the wireless channel may lead to the possibility of changes to the information, causing the occurrence of errors. The occurrence of errors is proportional to the amount of channel fading. Each wireless channel possesses an individual and independent fading that is different from the other channels. The impact of these different kinds of fading can be minimized using more than one independent path for the transfer of the same information. This method is implemented by using a number of transmitter antennas in Multiple -input multiple-output (MIMO) systems [1] [2]. A MIMO system represents a significant evolution in wireless communications [3]. Where such a system offers redundancy through the multiple independent channels, which are created between the transmitting and the receiving antennas of the system. Therefore, this system is introduced to enhance the performance of the wireless communication system, which provides robustness and increased reliability by overcoming channel fading with the use of multiple antennas. The MIMO systems have shown significant improvements in terms of the coverage and the data throughput without the need for additional transmission power or bandwidth expansion.

MIMO system can provide both spatial diversity and spatial multiplexing gains. It is important to note that all the gains provided by this scheme may not be achieved simultaneously. Instead, there is a trade-off between them. Although transmit diversity is clearly advantageous on a cellular base station, due to some practical determinants, the multiplicity of antennas in mobile devices is impractical. The solution to this problem is to encourage cooperation between mobile devices to make an array of antennas, that array can form a virtual transmit diversity [4]. This diversity technique is generated by transmitting many copies of the same information to different locations. As such these antennas are employed as a sender and relay in the network.

## 2. COOPERATIVE COMMUNICATION

In cooperative wireless communication the wireless agents, known as users have a single-antenna that can reap some of the benefits of the MIMO system which provides spatial diversity [5]. Cooperation in ad hoc networks (a network consisting of independent users) can save limited network resources, such as energy. It can also increase the reliability and the quality of service which is measured at the physical layer by bit error rates, or outage probability. Thus, it can extend the coverage area range and data throughput. Cooperative communication may be entitled to more benefits such as but not limited to (improved energy/power efficiency, increasing the attainable system throughput, cell-edge coverage extension, guaranteeing a given QoS, network deployment, throughput, improvement in data rate, mobility) [6].

In a cooperative communication system, it is assumed that each wireless user transmits data and acts as a cooperative agent for another user. Specifically, each user transmits both user's own bits as well as some information to the user's partner as shown in Figure 1. The findings have shown that the spectral efficiency of each user has improved by employing the cooperative technique. This is because the channel code rates have increased due to cooperation diversity.

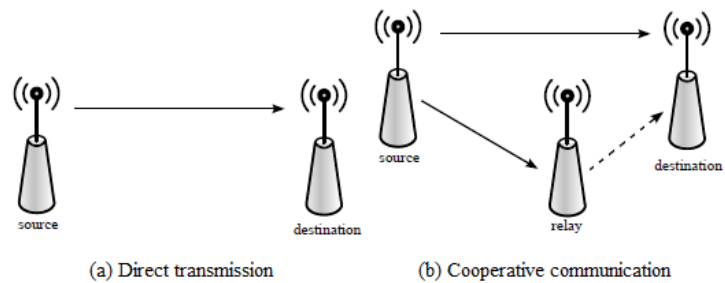

Figure 1. shows (a) direct transmission, (b) Cooperative transmission.

### 2.1 Amplify-and-Forward (AF)

In the wireless cooperative communication, Amplify and Forward (AF) is the simplest cooperative technique. In this technique, the signal received by the relay suffers from attenuation. Therefore, the noisy copy of the original signal needs amplification before sending it again by the relay, the application of the AF cooperative method is less difficult than the alternative cooperation methods where decoding and decision are not needed at the relay in the AF method [7].Meanwhile, the noise in the signal will be amplified too. At the destination, the transmitted signal by the user and its partner will be combined as shown in Figure 2. The last decision on the transmitted signal will is made at the destination. Although the AF technique amplifies the noise, the destination makes better decisions on the detection of information, because it receives two copies of information that were transmitted over independently fading channels. Laneman and Wornell [8] present and analyze the AF technique. Where to achieve the second order of diversity at least two users must be utilized. It was also shown that SNR with high values, the AF technique achieves the best possible BER performance. The use of inter-user channel characteristic is proposed because the destination is estimated to perform optimal decoding. This gain in the characteristics coefficient is very useful, as an indication that the relay is used to amplify the signal. Before transmission, this gain needs to be calculated. In the 1$^{st}$ time slot, relay will receive a copy of the signal that the source was transmitting. In the 2$^{nd}$ time slot, the relay will amplify the received signal according to the amplification factor that can be calculated as [8]

$$\beta = \sqrt{\frac{E_b}{E_b|h_{SR}|^2 + N_0}} \qquad (1)$$

Where $E_b$ is the energy of the transmitted signal, $h_{SR}$ is the channel fading coefficient between source-relay and the $N_0$ is the variance of the Additive White Gaussian Noise (AWGN) during the transmission in the uplink channel. The relay node uses the amplification factor $\beta$, and sends the amplified source signal to the destination node.

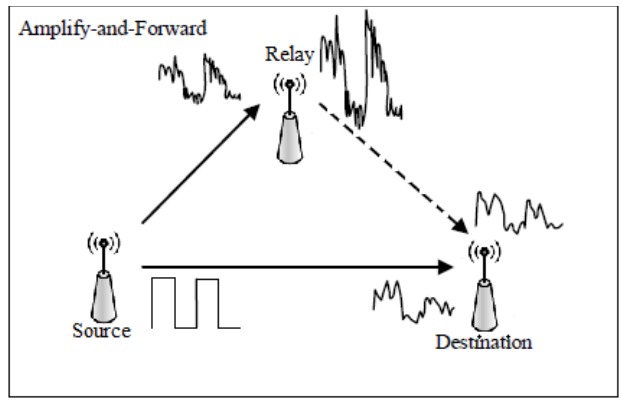

Figure 2. shows the (AF) Cooperative Technique.

## 2.2 Convolutional Coding

These codes are one of the powerful, effective and widely used in many applications. When a trivial corrupted data may become harmful or not used, then convolutional encoding can be implemented and transmitted data obtained with high accuracy [9]. A convolutional coding convolutes the data before the transmission by adding more bits. The goal of utilizing this technique is to minimize the probability of errors in transmission over the communication channel which is affected by fading conditions. The input to the convolutional encoder is a binary stream, the k symbol represents the number of input bits which is shifted into the shift register. Each k has a number of output bits that denoted as nbits. One more parameter called L that represent the constraint length of the code for the shift register. The convolutional code generator used is (31, 27) octal as shown below in Figure 3. Our system parameters take the values of k = 1, L = 5, n = 2 and the code rate (R = k/n) which equal to 1/2. The information encoding is done when passing a row of this matrix data through the encoder. The outputs n1 and n2 are calculated according to [g1]; [g2], that their input from Q1;Q2;Q3;Q4 and Q5. Here, the two vectors generators are [g1] = [11001],[g2] = [10111]. AQPSK modulation will be utilized to modulate the encoded bit stream.

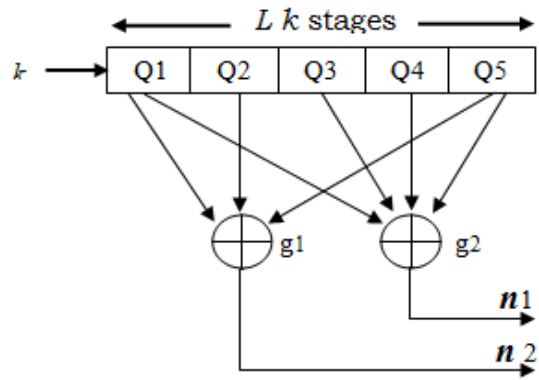

Figure 3. Convolutional encoder with a code rate = ½.

## 3. SYSTEM MODEL

The System topology considered in this paper is depicted in Figure6. It is worth noting that this research could also be applied to other applications, such as home entertainment, wireless body area network (WBAN), Internet of Things (IoT), and generic sensor networks. Each node is equipped with one TX antenna and one RX antennas (in a typical scenario) [10]. The transceiver and receiver design as shown in figures 4 and 5 respectively, bits in the source are fed to a convolutional encoder with rate=1/2, and afterward to be mapped to QPSK symbols and transmitted at the first time slot to both destination and relay nodes.

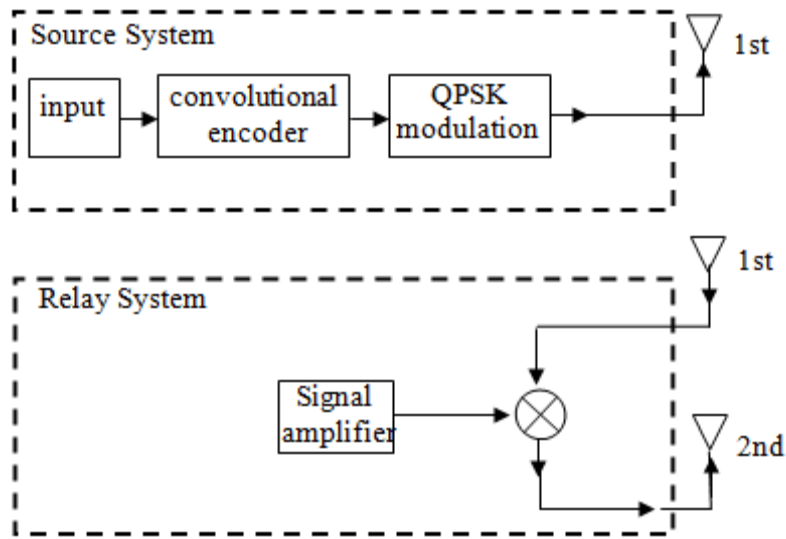

Figure 4. Transceiver block diagram for source and its partner.

The relay will resend the received signal to the destination at the second time slot after amplifying it. Figure 5. Demonstrate the destination node that will receive a multicopy for the same signal and combine these copies in the combiner to be demaped in the QPSK demodulator then will decode it by the convolutional decoder.

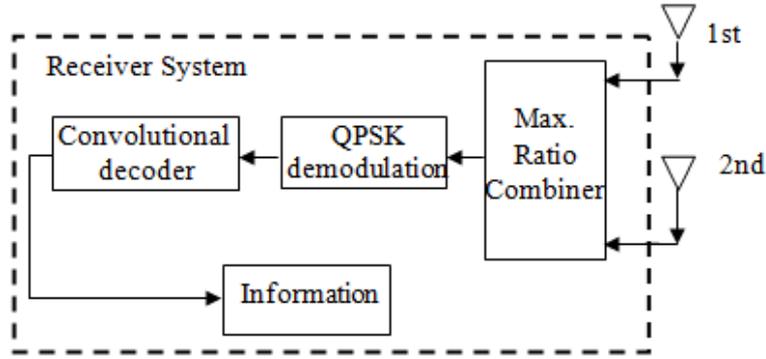

Figure 5. Block diagram for receiver node.

## 3.1 Relay location

In this paper, there are three cases we have applied in the system to be analyzed. These cases depend on the deployment (location) of a relay device as in Figure 6:

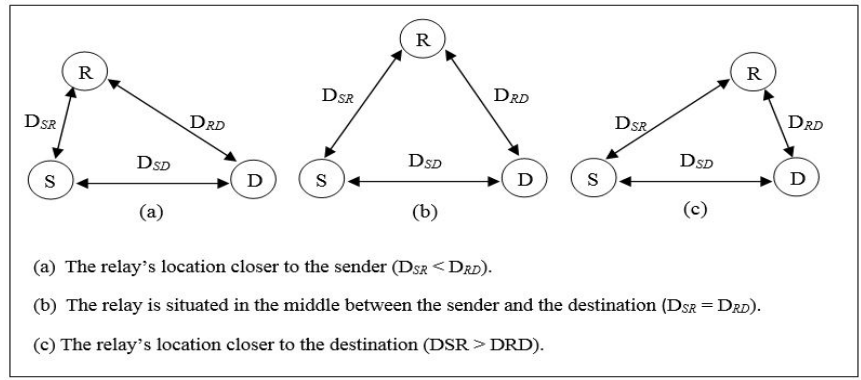

Figure 6. shows the relay at various positions.

Where in Figure6 (b) the arrangement is illustrates the first case, where the relay is at an equal distance between it and between both source and destination.

## 3.2 Theoretical Analysis

In large scale fading the path loss is estimated, where a path loss is directly proportional to the distance as shown in the equation [11]:

$$PL(A,B) = Q/(d_{A,B})^\alpha \qquad (2)$$

Where $Q$ is a constant that depends on the environment, $d_{A,B}$ denote the distance between nodes A and B, and α is the path loss exponent equal to four (obstructed building) [12]. Assuming a unit path loss between S and D, the received energy at the relay can be related to the received energy at the destination according to

$$E(s,r) = \frac{PL(s,r)}{PL(s,d)} E(s,d) = \left(\frac{d_{s,d}}{d_{s,r}}\right)^4 E(s,d) = G_{sr} E(s,d) \qquad (3)$$

Where $E(s, d)$ is the average transmitted (and thus received) energy between the S and D channel link. Therefore, the power gain (or geometrical gain) experienced by the inter-user channel with respect to the uplink between source and destination is:

$$G_{sr} = (d_{sd}/d_{sr})^4 \qquad (4)$$

The received signal at the relay node during the first transmission can be written as:

$$y_{sr} = \sqrt{G_{sr}} h_{sr} x_s + n_{sr} \qquad (5)$$

The power-gain at the uplink between relay and destination with respect to the uplink between source and destination can be computed as:

$$G_{rd} = (d_{sd}/d_{rd})^4 \qquad (6)$$

The symbols are transmitted from the relay node is given by:

$$y_{rd} = \sqrt{G_{rd}} h_{rd} x_r + n_{rd} \qquad (7)$$

The power-gain at the uplink between source and destination with respect to itself is unity:

$$G_{sd} = 1 \qquad (8)$$

The received signal at the destination node can be expressed as:

$$y_{sd} = h_{sd} x_s + n_{sd} \qquad (9)$$

## 3.3 Channel Model

Due to line-of-sight (LOS) propagation, the strongest propagation component of MIMO channel corresponds to the deterministic component (also referred to as specular components). On the other hand, all the other components are random components (due to NLOS also referred to as scattering components) [13]. The broadcast channel distribution has been following the Rayleigh channel distribution which is Gaussian distribution with a variance of σ² and zero mean. That means there is no component of LOS (K= 0):

$$\sigma = \sqrt{\frac{1}{K+1}} \qquad (10)$$

On the other hand, when there is any component of LOS (For K > 0) the broadcast channel distribution has been following the Gaussian distribution with a variance of σ² and mean of q or Rician distribution when K increases as:

$$q = \sqrt{\frac{K}{K+1}}, \qquad \sigma = \sqrt{\frac{1}{K+1}} \qquad (11)$$

Therefore, in this work, channel matrix of MIMO system is tending to be described as [14]:

$$H = \sqrt{\frac{K}{K+1}} H_d + \sqrt{\frac{1}{K+1}} H_r \qquad (12)$$

Where $H_d$ represents the component of the normalized deterministic channel matrix, while $H_r$ represents the component of random channel matrix, with $||H_d||^2 = N_T M$, E$\{|[H_r]_{i,j}|^2\}= 1$, $i = 1:N_T, j = 1: M$[14]. While K is known as a factor of the Rician channel which is the relation between the component of the specular power $c^2$ and the component of scattering power $2\sigma^2$, displayed as [13]:\

$$K = \frac{||H_d||^2}{E\{|[H_r]_{i,j}|^2\}} = \frac{c^2}{2\sigma^2} \qquad (13)$$

## 4. TOPOLOGY MODELS ANALYSIS

### 4.1 Equilateral Triangle Topology

The equilateral triangle topology as in Figure 6 (b), the distance between the source, destination and relay is equal. Therefore, when substituting in (4), the power gain will be equal at each node as follows:

$$G_{sr} = G_{rd} = G_{sd} = 1 \qquad (14)$$

### 4.2 Isosceles Triangle Topology

Isosceles triangle topology which represented in Figure 6 as (a) and (c) the arrangement is illustrated where the relay is closer to the sender than the destination and closer to the destination than the sender, respectively. In this topology relay location will not be random but is governed by a mathematical relationship. In first case as shown in figure 7 (a) when $d_{sd} = d_{sr} = d$, and $d_{rd} < d$

$$d_{rd} = d\sqrt{2(1 - \cos\theta)}, \quad \theta \in (0, \pi/3)$$

If $\theta = \pi/4$ and $d = 1$, $d_{rd}$ will equal to (0.765367), then

$$G_{rd} = (d_{sd}/d_{rd})^4 = (1/0.765367)^4, \qquad G_{sr} = G_{sd} = 1$$

In second case as shown in figure 7 (b) when $d_{sd} = d_{rd} = d$, and $d_{sr} < d$

$$d_{sr} = d\sqrt{2(1 - \cos\varphi)}, \quad \varphi \in (0, \pi/3)$$

If $\varphi = \pi/4$ and $d = 1$, $d_{sr}$ will equal to (0.765367), then

$$G_{sr} = (d_{sd}/d_{sr})^4 = (1/0.765367)^4, \qquad G_{rd} = G_{sd} = 1$$

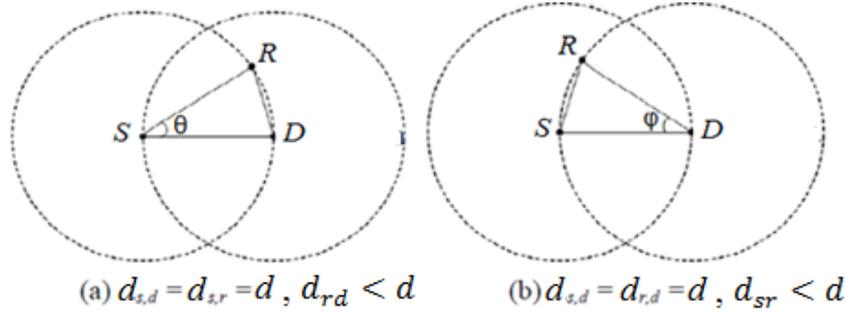

Figure 7. illustrate the Isosceles triangle topology.

## 4.3 Linear Topology

In this topology the relay $R$ moving in straight path along the $SD$ distance between the source and destination, assuming $d_{sd}$ is a fixed value, $d_{sr} = \rho \cdot d_{sd}$. When $\rho \in (0,1)$, $d_{rd} = (1 - \rho) d_{sd}$, where $\rho$ represents the amount of the relay movement. The process of determining the optimal location of relay is as follows.

$$G_{sr} = (d_{sd}/\rho\, d_{sd})^4 = (1/\rho)^4$$
$$G_{rd} = (d_{sd}/(1-\rho)\, d_{sd})^4 = (1/(1-\rho))^4$$

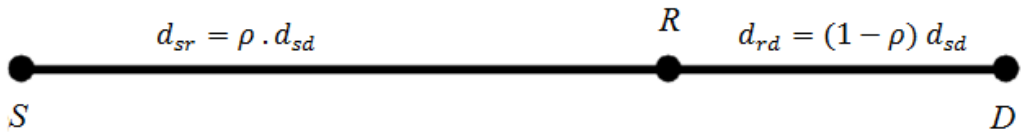

Figure 8. illustrate the Linear topology.

## 4.4 Scalene Triangle Topology

In scalene triangle topology, we need to calculate the gain when the relay moving, to do so, the $f$ which represents the distance between $R$ and $d_{sd}$ must be calculated first, and it will take different values as follows:

**A.** In Figure 9 (b). if $d_{sr} = d_{rd} = d_{sd} = a$ then

$$f = \sqrt{a^2 - (a/2)^2} \tag{15}$$

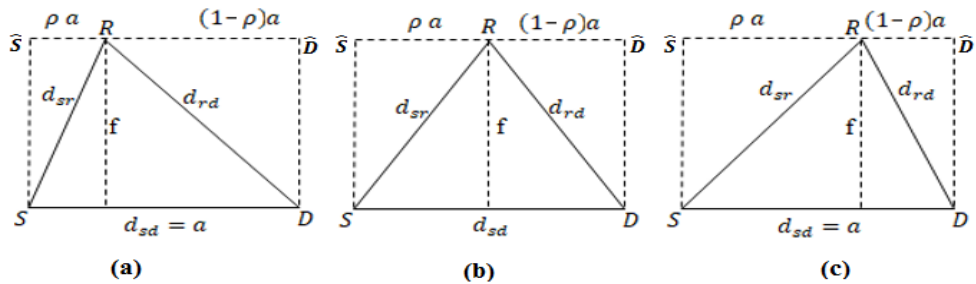

Figure 9. illustrate the Scalene triangle topology.

In figure 9 the $(\hat{S}\hat{D})$ line represents a parallel reflection of the $(SD)$ line, where $\hat{S}R$ is $d_{\hat{S}R} = \rho\alpha$, $R\hat{D}$ is $d_{R\hat{D}} = (1 - \rho)\alpha$. Now if $R$ moves along the $(\hat{S}\hat{D})$ line, closer to the $\hat{S}$, then the equilateral triangle in in figure 9 (b) became Scalene Triangle as in Figure 9 (a), that mean the distance $d_{\hat{S}R} < a/2$, then the $d_{sr}$ will calculated as follows:

$$d_{sr} = \sqrt{f^2 + (\rho a)^2} \qquad (16)$$

As well as $d_{rs}$ can be calculated:

$$d_{rd} = \sqrt{f^2 + [(1-\rho)a]^2} \qquad (17)$$

If $d_{sd} = 1 = a$, by substituting in (15) then $f = 0.866$, and if $\rho = 0.35$, then the $d_{sr}$ will be as in (16):

$$d_{sr} = \sqrt{(0.866)^2 + (0.35)^2} = 0.93$$

And $d_{rd}$ can be calculated as in (17):

$$d_{rd} = \sqrt{(0.866)^2 + (1-0.35)^2} = 1.08$$

The gain can obtain now as

$$G_{sr} = (d_{sd}/d_{sr})^4 = (1/0.93)^4 \ , G_{rd} = (d_{sd}/d_{rd})^4 = (1/1.08)^4$$

The same mathematical approach applies if $R$ moves along the parallel line of the $d_{sd}$ line, closer to the $\hat{D}$, which represented in figure 9 (c), then $d_{\hat{S}R} > a/2$. Again, if $d_{sd} = 1 = a$, then $f = 0.866$, and if $\rho = 0.65$, then the $d_{sr}$ will be as in (16) we will get $d_{sr} = 1.08$, and $d_{rd} = 0.93$. The gain will be as follows:

$$G_{sr} = (1/1.08)^4 \ , G_{rd} = (1/0.93)^4 \ , G_{sd} = 1$$

**B.** Here we will study the case when the distance between $\hat{S}\hat{D}$ line and $SD$ line, that distance represented by ($f$), and has a different value under the condition $f > 0$, where $f = 0$ that means the linear topology.

In Figure 10 (b). if $d_{sr} = d_{rd} \neq d_{sd}$, $d_{sd} = a = 1$, $f = a(3/4)$

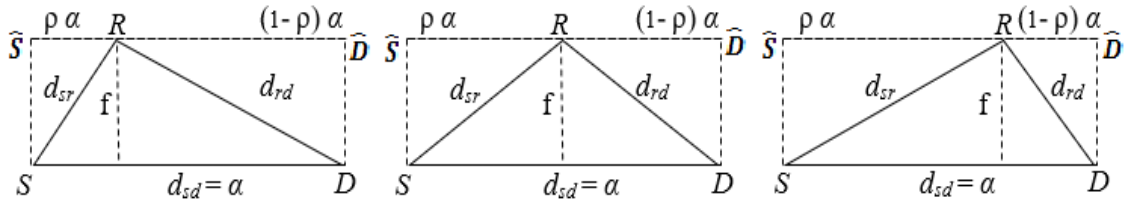

Figure 10. illustrate the scalene triangle topology.

If $R$ moves along the ($\hat{S}\hat{D}$) line, closer to the $\hat{S}$, then $d_{\hat{S}R} < a/2$, and if $\rho = 0.35$, then the $d_{sr}$ will calculated by substituting $f$ in (16):

$$d_{sr} = \sqrt{(0.75)^2 + (0.35)^2} = 0.82$$

We can say the same for $d_{rs}$ as (17):

$$d_{rd} = \sqrt{(0.75)^2 + [(1-0.35)a]^2} = 0.99$$

The gain can obtain now as:

$$G_{sr} = (1/0.82)^4 \ , G_{rd} = (1/0.99)^4 \ , \ G_{sd} = 1$$

Again, the mathematical module in equation (16) and (17) can be apply if *R* moves along the parallel line of the $d_{sd}$ line, closer to the $\widehat{D}$, when $d_{\widehat{S}R} > \alpha$ /2 and $\rho$=65 subsequently we will get:

$$G_{sr} = (1/0.99)^4 \ , G_{rd} = (1/0.82)^4, \ G_{sd} = 1$$

## 5. SIGNAL COMBINER

The Maximum Ratio Combiner (MRC) was adopted as a signal Combiner at the destination in this paper, the MRS achieves the best possible performance. Each input signal of the 1st and 2nd time slots in the distention will multiply with its corresponding conjugated channel gain [11].

$$r_d = \sum_{i=1}^{k} h^*_{id} u_{id} \qquad (18)$$

This study is using one relay system. Therefore, this equation can be rewritten as:

$$r_d = h^*_{sd} u_{sd} + h^*_{rd} u_{rd} \qquad (19)$$

When looking to this equation a little closer, it's easy to notice the disadvantage of this method in a multi-hop environment. The MRC only considers the last hop (i.e. the last channel) of a multi-hop link. So, the use of MRC is just recommended when an error correcting is employed in the system. This is accomplished in this study by using the convolution code.

## 6. SIMULATION RESULTS

The systems presented in this paper are simulated using Matlab codes. In this part, It was evaluated the signal-to-noise ratio (SNR) against the bit error rate (BER) as a scale of the system performance over Rayleigh fading channel a conventional channel and Rician fading channel as a realistic channel. For random Rayleigh and Rician fading channel case, the samples of these parameters are set up to 1000 with elements generated as zero-mean for Rayleigh fading channel while m-mean for Rician fading channel and unit-variance are independent and identically distributed (i.i.d) complex Gaussian random variables. QPSK signal constellation has been used as a broadcast modulation in all simulations and the results are averaged through several channel investigations.

Table 1 represents the summary of parameters used in our system. Figure11 shows the equilateral triangle topology BER performance of the cooperative system as presented in [15], and compared to our system that employs convolutional cooperative. It shows clearly the effect of harnessing the convolution code in such a transmission system which can improve the BER performance of the system.

Table 1. Shows the system parameters.

| parameter | |
|---|---|
| Number of relays | 1 |
| Source power allocation | 1/2 – 2/3 of total transmission power |
| Relay power allocation | 1/2 – 1/3 of total transmission power |
| Inter user channel type | Rayleigh and Rician |
| Rician channel factor (K) | 15 - 20 |
| SNR of inter user channel | 20 - 30 |
| Number of antennas for each node | 1 |

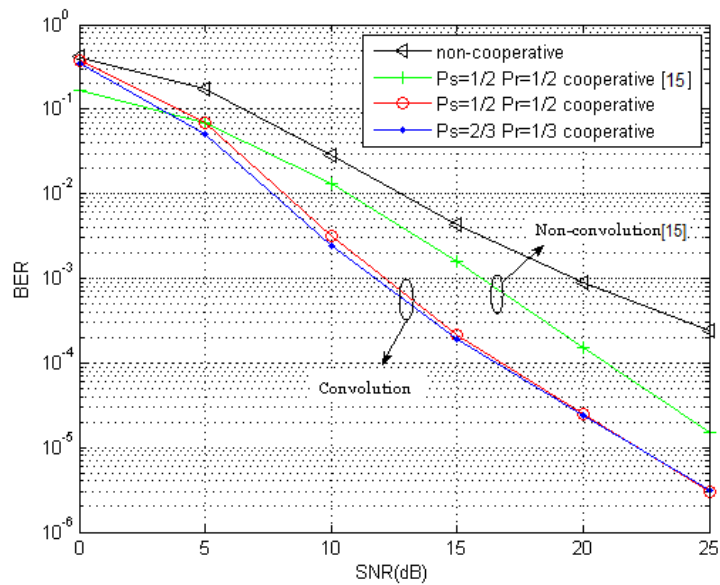

Figure 11. Convolutional code effects on the cooperative system performance, with $d_{sd}=d_{sr}=d_{rd}=1$ and various allocated power, Source power (Ps) and Relay power (Pr).

More specifically, to achieve a BER of about $10^{-4}$ the required SNR for the convolution coded system is less by 5 dB than the non-convolutional coded system.

Figure 12 shows the BER performance of the convolutional cooperative system when communicating over isosceles triangle topology with different relay positions. Figure 12 compared to the same topology in [15], where the non-convolutional system offers a better performance from (0-5) dB, while the SNR increase the convolutional system shows an outperformance from the non-convolutional system. Also, it is showing that when both equal and unequal power allocated for source-relay, the relay closer to distance achieved better BER performance.

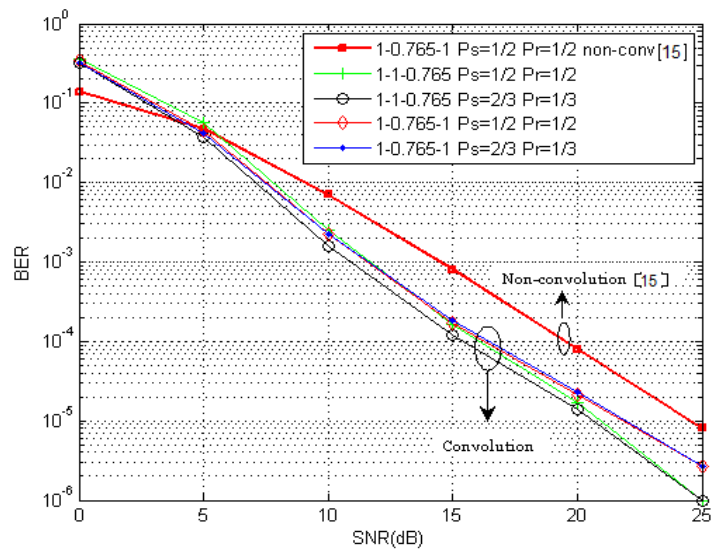

Figure 12. BER performance of isosceles triangle topology with different $d_{sd}$, $d_{sr}$, $d_{rd}$.

This simulation result demonstrates that as the distance between the destination and relay increase BER performance become worse, and when the relay is far from the source the BER will be better regardless the allocated power, where to achieve BER of about $10^{-5}$, where the required SNR is about 21dB, and the BER $10^{-6}$ can be achieved with SNR 25dB.

Figure 13 shows the comparison of the BER performance of the system when communicates over the linear topology, again the result compared to that in [15]. Moreover, the convolutional system gets the better BER performance. The simulation shows that when relay located at the middle distance between source and destination it can achieve the best BER performance, which is about $10^{-6}$ at 22 dB, in contrast with the previous topologies. In linear topology the equal power allocation offers slightly better BER performance, while in others unequal power allocation did the best performance.

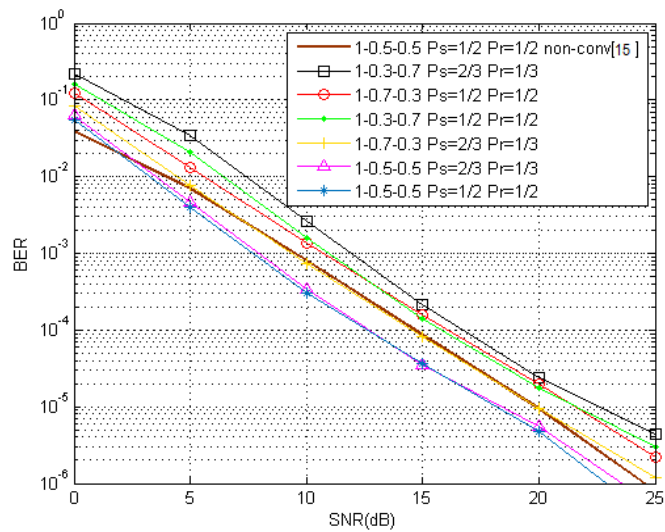

Figure 13. BER performance of linear topology with different $d_{sd}$, $d_{sr}$, $d_{rd}$.

Figure 14(a-b) simulates the system BER performance with different power allocation over scalene triangle topology (A) and (B) respectively.

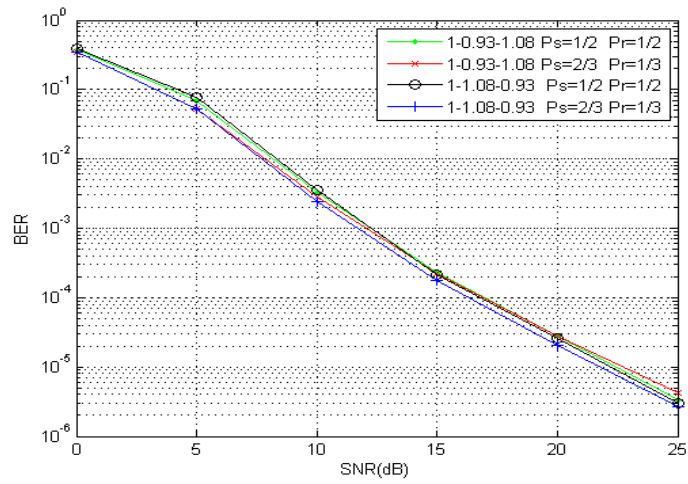

Figure 14(a). BER performance of scalene triangle topology (A) with different $d_{sd}$, $d_{sr}$, $d_{rd}$.

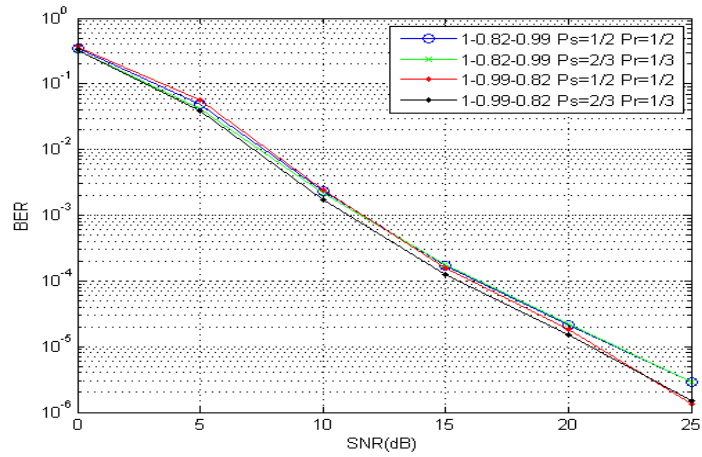

Figure 14(b). BER performance of scalene triangle topology (B) with different $d_{sd}$, $d_{sr}$, $d_{rd}$.

In Figures 14(a-b) we barely notice a performance difference between the relay positions in each topology separately.

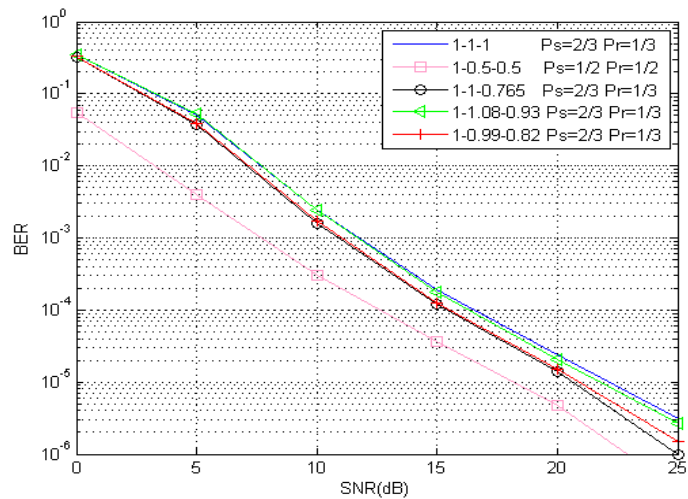

Figure 15. The comparison of performance among topologies over Rayleigh channel.

International Journal of Computer Networks & Communications (IJCNC) Vol.11, No.3, May 2019

The comparison in simulation of BER among equilateral triangle topology, isosceles-triangle topology (b) (π/4, Rsr =0.765), linear topology (ρ =0.5), scalene triangle topology (A) (Rsr=0.81) and (B) (Rsr=0.79) is shown in Figure 15. We can see that the BER performance of linear topology (ρ =0.5) is the best. Isosceles triangle topology (b) (π/4, Rrd =0.765) has the same BER performance of scalene triangle topology (B) (Rsr=0.99, Rrd=0.82) and is better than that of equilateral triangle topology which is equal to scalene triangle topology (A) (Rsr=1.08, Rrd=0.93).

As shown in Figure 16, the inter-user (source-relay) channel was simulated as Rician fading channel. In this case we can see that the a Rician fading channel has a significant effect on the BER performance overall topologies.

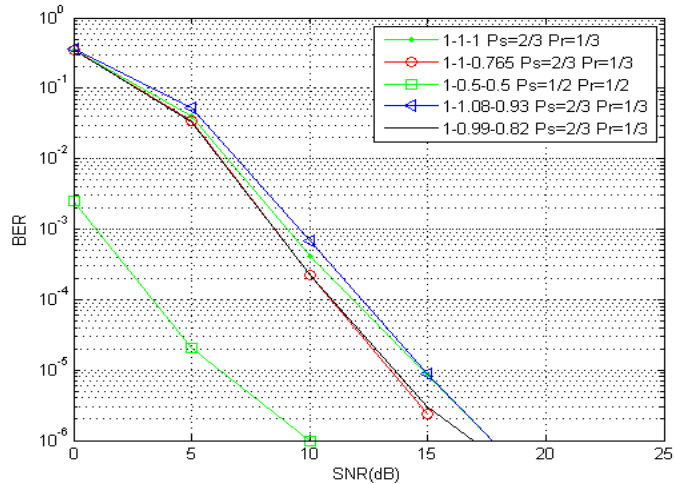

Figure 16. The comparison of performance among topologies over Rician inter-user channel.

In comparison with Figure 15, Figure 16 shows that the system over Rician channel offers better BER performance than over Rayleigh channel. The result shows that the BER performance of the AF method is improved as the quality of the inter-user channel improves. Because in this method, the signal received by each relay is a noise signal, it suffers from attenuation. Therefore, the noisy version of the original signal needs to be amplified before it can be sent again by the relay. In doing so, the noise in the signal is also amplified. From this analysis it can be seen that the line-of-sight (LOS) environment propagation is supporting the AF cooperation method.

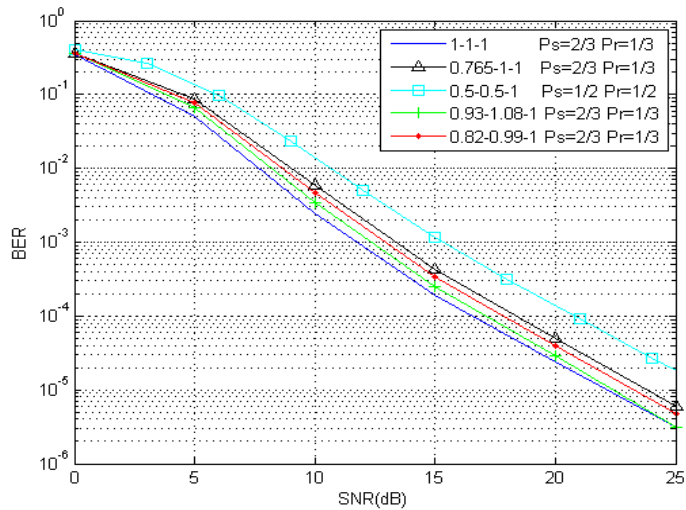

Figure 17. The comparison of performance among topologies over Rayleigh channel when source-relay act as relay-source.

.In Figure 17, the role of source-relay in the system was replaced by relay-source, which means the relay takes the source's position and vice versa. The system performance overall topologies was simulated. The results show that the equilateral triangle outperforms all the other topologies. Furthermore, the equilateral triangle topology offers a very steady performance no matter how the source-relay replaces its positions. The same system with the inter-user Rician channel was simulated in Figure 18. We can see that the scalene a topology gets the best BER performance.

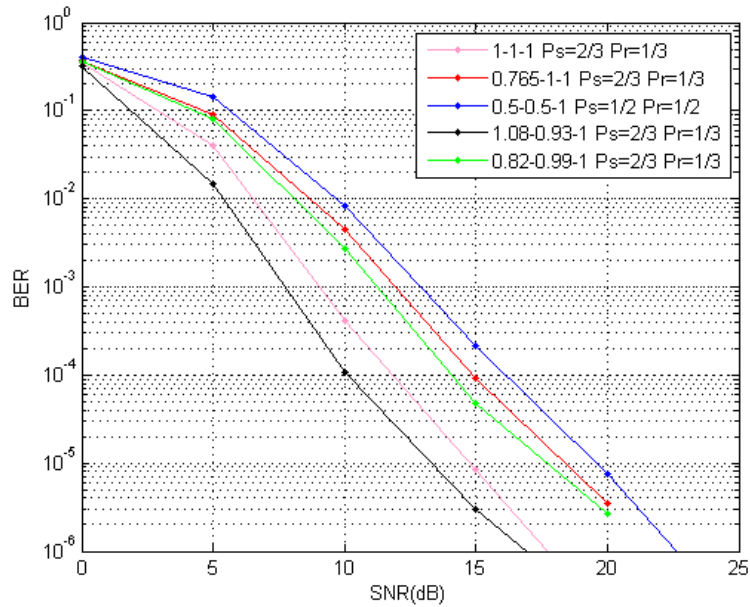

Figure 18. The comparison of performance among topologies when source-relay acts as relay-source and its channel is Rician channel.

## 7. CONCLUSION

In this paper, the performance of relayed communication by using the convolutional code with different power allocation has been evaluated in different topologies over Rayleigh and Rician fading channel. Also, we investigated these topologies to cover all the possible positions of the relay when moving to a different location. It is shown that cooperative system provides much better BER performance as compared with the data in which no cooperation used. Furthermore, convolutional coding has a superior impact on enhancing the performance of cooperative communication. By the simulation results, it is observed that the linear topology with equal power allocation offers the best performance. Meanwhile, the other topologies show that varying the relay position far from the destination will disturb the ability of the system to correct errors. Also, it was observed that when the source-relay replaced positions the equilateral triangle topology with non-equal power allocation shows an outperformance as well as a stable BER performance over the other topologies. We analyzed the performance of the system in the LOS environment (Rician fading channel) and in the NLOS environment (Rayleigh fading channel). When the inter-user channel simulated as the Rician fading channel, it is shows much better BER performance than the Rayleigh fading channel.

**REFERENCES**

[1]  E. Biglieri, R. Calderbank, A. Constantinides, A. Goldsmith, A. Paulraj and H.V. Poor, MIMO Wireless Communications, Cambridge University Press 2007.

[2]  G. J. Foschini, Layered Space-Time Architecture for Wireless Communication in a Fading Environment When Using Multi- Element Antennas, Bell Laboratories Technical Journal 41–59, October 1997.https://doi.org/10.1002/bltj.2015.

[3]  Gong, D., Zhao, M., Yang, Y, A multi-channel cooperative MIMO MAC protocol for clustered wireless sensor networks, Journal of Parallel and Distributed Computing, 74 (2014), pp. 3098-3114https://doi.org/10.1016/j.jpdc.2014.07.012.

[4]  S. Aneja, S. Sharma, A novel cooperative communication sysem based on multilevel convolutional codes, Wireless Personal Communications, vol. 95, no. 4, pp. 3539-3556, 2017 https://doi.org/10.1007/s11277-017-4011-z.

[5]  S. M. Alamouti, A simple transmit diversity technique for wireless communications, IEEE J. Select. Areas Commun., vol. 16, pp. 1451–1458, October 1998. https://doi.org/10.1109/49.730453.

[6]  A. Abraham, N. Madhu. Cooperative Communication For 5G Networks: A Green Communication Based Survey, International Journal of Information Science and Computing: 4(2): December 2017: p. 65-78.http://dx.doi.org/10.5958/2454-9533.2017.00007.2.

[7]  Z. Mo, W. Su, and J. D. Matyjas, Amplify and forward relaying protocol design with optimum power and time allocation. In MILCOM 2016 - 2016 IEEE Military Communications Conference, pages 412–417, Nov 2016 https://doi.org/10.1109/MILCOM.2016.7795362.

[8]  J. N. Laneman and G. W. Wornell, Cooperative diversity in wireless networks: Efficient protocols and outage behavior, IEEE Trans. Inform. Theory, vol. 50, pp. 3062-3080, 2004. https://doi.org/10.1109/TIT.2004.838089.

[9]  S. Madan La, S. Vivek Kumar, Effects of Code Rate and Constraint Length on Performance of Convolutional Code, International Journal of Information, Communication and Computing Technology, 2016, Volume: 4, Issue: 1.

[10] L.C.Tran, X.Huang, A.Mertins, F. Safaei, Comprehensive Performance Analysis Of Fully Cooperative Communication in WBANs, IEEE Journals & Magazines , Volume: 4 ,2016.https://doi.org/10.1109/ACCESS.2016.2637568.

[11] H. Ochiai, P. Mitran, and V. Tarokh, Design and analysis of collaborative diversity protocols for wireless sensor networks, in Vehicular Technology Conference, 2004. VTC2004-Fall. 2004 IEEE 60th, vol. 7, pp. 4645–4649, IEEE, 2004.https://doi.org/10.1109/VETECF.2004.1404971.

[12] A. Meier and J. Thompson, Cooperative diversity in wireless networks, in 3G and Beyond, 2005 6th IEE International Conference on, pp. 1–5, IET, 2005.https://ieeexplore.ieee.org/document/4222750.

[13] Y. S. Cho, J. Kim, W. Y. Yang, C-G. Kang, MIMO-OFDM Wireless Communications with MATLAB, 2010, https://doi.org/10.1002/9780470825631 .

[14] X Shi, C Siriteanu, S Yoshizawa and Yoshikazu Miyanaga, 2012. Realistic Rician Fading Channel based Optimal Linear MIMO Precoding Evaluation, in Communications Control and Signal Processing (ISCCSP), 5th International Symposium.https://doi.org/10.1109/ISCCSP.2012.6217780

[15] Lei Xu, Hong-Wei Zhang,  Optimum Relay Location in Cooperative Communication Networks with Single AF Relay,  in Int. J. Communications, Network and System Sciences, 2011, 4, 147-151 DOI: 10.4236/ijcns.2011.43018.